\documentclass{amsart} 
\usepackage[all]{xy} 
 
\font \fraktur     = eufm10 scaled \magstephalf 
 
\newtheorem{thm}{Theorem}[section] 
\newtheorem{prop}[thm]{Proposition} 
\newtheorem{cor}[thm]{Corollary} 
 
\theoremstyle{definition} 
 
\newtheorem{ex}[thm]{Example} 
\theoremstyle{remark} 
\newtheorem{rem}[thm]{Remark}

\def\K{{\mathcal K}}         
\def\T{{\mathcal T}}         
\def\M{{\mathcal M}}         
\def\P{{\mathcal P}}         
\def\Lie{{\mathcal L}}       
 
\font \fraktur=eufm10 scaled \magstephalf 
\def \frak#1{\hbox{\fraktur #1}} 
\def \sllie{{\frak s}{\frak l}} 
 
\def \gg{{\frak g}} 
\def \hh{{\frak h}} 
 
\def\R{\mbox{$\mathbb R$}} 
\def\C{\mbox{$\mathbb C$}} 
 
\def\Z{\mbox{$\mathbb Z$}}

\def\pb#1#2{\left\{#1,#2\right\}}                
\def\Pb{\pb{\cdot}{\cdot}} 
 
\def\pbt#1#2{\left\{#1,#2\right\}_{\T}}          
\def\Pbt{\pbt{\cdot}{\cdot}} 
\def\pbk#1#2{\left\{#1,#2\right\}_{\K}}          

\def\tr{\mathop{\hbox{\rm tr}}\nolimits} 
\def\rank{\mathop{\hbox{\rm rank}}\nolimits} 
 
\def\set#1{\left\{#1\right\}}

\newcommand{\comment}[1]{} 

\begin{document} 
\title[From Toda to Volterra]{From the Toda Lattice to the Volterra
  Lattice and back} 
\author{Pantelis A. Damianou} 
 \address{Department of Mathematics and Statistics, The University of 
   Cyprus, P.O.~Box 537, Nicosia, Cyprus} 
 \email{damianou@ucy.ac.cy} 
\author{Rui Loja Fernandes} 
 \address{Departamento de Matem\'atica, Instituto Superior T\'ecnico, 
   1049-001 Lisboa, Portugal} 
 \email{rfern@math.ist.utl.pt} 
 \thanks{The second author was supported in part by FCT-Portugal 
 through program POCTI and grant POCTI/1999/MAT/33081.} 
 
\renewcommand{\subjclassname}{\textup{2000} Mathematics Subject 
     Classification} 
\begin{abstract} 
We discuss the relationship between the multiple Hamiltonian 
structures of the generalized Toda lattices and that of the 
generalized Volterra lattices. 
\end{abstract} 
\maketitle 
\tableofcontents 
 
\section{Introduction} 
\label{intro} 
 
The Volterra lattice is the system of o.d.e.'s 
\begin{equation} 
\label{eq:KM} 
  \dot a_i=a_i(a_{i-1}-a_{i+1})\qquad (i=1,\dots,n), 
\end{equation} 
where $a_{n+1}=a_0=0$. This system was first studied by Kac, 
van-Moerbeke and Moser in two foundational papers (\cite{Kac,Mos}) 
for the modern theory of integrable systems. They showed, for 
example, that this system arises as a finite dimensional 
approximation of the famous KdV equation. In this paper we shall 
refer to system (\ref{eq:KM}) as the \emph{$A_{n}$-Volterra system}.

Another well known discretization of the KdV equation is the Toda 
lattice \cite{Tod}. The Toda system can be written in the form: 
\begin{equation} 
\label{eq:Toda} 
\left\{ 
\begin{array}{ll} 
  \dot a_i=a_i(b_i-b_{i+1}),& \\ &\qquad (i=1,\dots,n) \\ \dot 
  b_i=a_{i-1}-a_i,& 
\end{array} 
\right. 
\end{equation} 
with $a_{0}=b_{n+1}=0$. We shall 
refer to this system as the \emph{$A_{n-1}$-Toda lattice}. 
Generalizations of both equations to root systems other than $A_n$
were obtained by Bogoyavlenskij in \cite{Bog1,Bog2}. 

For the $A_n$-Toda lattice, quadratic and cubic brackets were
constructed, respectively, by Adler in \cite{adler} and by Kupershmidt
in \cite{kup}. Multiple Hamiltonian structures for these systems were
introduced by the first author (\cite{Dam1}) in Flaschka coordinates
and by the second author (\cite{Fer1}) in natural $(q,p)$
coordinates. The analogous results for $B_n$-Toda were computed in
\cite{Dam3} in Flaschka coordinates, and in \cite{DN} in $(q,p)$
coordinates. The $C_n$ case is in \cite{Dam4} in Flaschka coordinates,
and in \cite{DN} in natural $(q,p)$ coordinates. Finally, the $D_n$ case can be
found in \cite{Dam5}.  The construction of these multiple Hamiltonian
structures for the exceptional root systems is an open problem.

For the Volterra system, multiple Hamiltonian structures 
were constructed recently by Kouzaris (\cite{Kou}) for $B_n$, $C_n$ 
and $D_n$ systems, which generalize the $A_n$-case. For the $A_n$ case,
it was proved in \cite{Fer2} that the (periodic)  
Volterra lattice is an algebraic completely integrable (a.c.i.) 
system.  In \cite{Fer3}, the hyperelliptic systems were introduced 
similar to the even and odd Mumford systems (see \cite{Mum1,Van1}), 
and the following link was established: there is a commutative diagram 
\smallskip 
\[ 
\newdir{ (}{{}*!/-5pt/@_{(}} 
\xymatrix{ 
\T\ar[r]^{\Phi}&\M \\ 
\K\ar[r]_{\Phi}\ar@{ (->}[u]& \P\ar@{ (->}[u]} 
\] 
in which $\M,\,\P,\,\T,\,\K$ are (in that order) the phase spaces 
of the (even) Mumford system, the hyperelliptic Prym system (odd or 
even), the periodic $A_n$-Toda lattice and the periodic 
$A_n$-Volterra system.  The vertical arrows are natural inclusion 
maps exhibiting for both spaces the subspace as fixed points 
varieties and the horizontal arrows are injective maps that map 
every fiber of the momentum map on the left injectively into (but 
not onto) a fiber of the momentum map on the right. In addition, 
Poisson structures were constructed so that this diagram has a 
meaning in the Poisson category. These give a precise geometric 
description, in the $A_n$ case, of the well known  connection 
between the Toda and Volterra systems (see also \cite{Ves,Vol}). 
 
In this paper we initiate the study of the relationship between the 
Toda and Volterra systems, for root systems other than $A_n$.  Here 
we shall restrict our attention mainly to the $B_n$ case, for which we 
give a complete description.  The analysis for this case is 
simplified since we can view this system as a subsystem of the 
$A_{2n}$-system. We exhibit the phase space of the $B_n$-system as 
a fixed point set of a finite group action by Poisson 
automorphisms.  We prove a general result which allow us to reduce 
the Poisson structures in such situations, and hence we can relate 
the multiple Hamiltonian structures of the Volterra and Toda 
systems. 
 
The plan of this paper is as follows: In Section \ref{Poisson}, we
discuss several conditions under which the fixed point set of a
Poisson action inherits a Poisson bracket, generalizing the Poisson
involution theorem (see \cite{Fer3,Xu}). In Section \ref{Toda}, we
recall the definition of the Toda systems, and give the multiple
Hamiltonian structure for the $B_n$ system. These were discovered in
\cite{Dam3} (see also \cite{DN}) and we use here a new approach based
on the Poisson involution theorem to deduce them.  In Section
\ref{Volterra}, we first recall the construction of the Volterra
system and its generalizations for any root system associated with a
simple Lie algebra.  Then, we give the multiple Hamiltonian structure
for the $B_n$ (and $C_n$) cases and, by applying a symmetry approach,
we explain the relation between the Poisson structures for Volterra
and Toda lattices. In the final section, we describe the connections
we have been able to find between the two systems, including a
generalization of Moser's recipe (see \cite{Mos}) to pass from the
Volterra to the Toda lattices.
 
%
\section{Fixed point sets of Poisson actions}       
%
\label{Poisson} 

We shall see that the relation between the Toda and Volterra systems 
relies on special symmetries of the phase spaces. In this section, we 
study conditions under which the fixed point set of a Poisson action 
inherits a Poisson bracket. 
 
Although we will be interested mainly in finite symmetries, we give 
the following general result: 
 
\begin{thm} 
\label{thm:invol:poisson} 
  Suppose that $(M,\Pb)$ is a Poisson manifold, and $G$  a compact 
  group acting on $M$ by Poisson automorphisms. Let $N=M^G$ be the 
  submanifold of $M$ consisting of the fixed points of the action and 
  let $\iota:N\hookrightarrow M$ be the inclusion. Then $N$ carries a 
  (unique) Poisson structure $\Pb_N$ such that 
  \begin{equation} 
    \label{eq:invol:poisson} 
    \imath^*\pb{F_1}{F_2}=\pb{\imath^*F_1}{\imath^*F_2}_N 
  \end{equation} 
  for all $G$-invariant functions $F_1,F_2\in C^\infty(M)$. 
\end{thm} 

\proof 
For $f_1,f_2\in C^\infty(N)$ we choose $F_1,F_2\in C^\infty(M)$ such that 
$f_i=\imath^*F_i$. We may assume that $F_1$ and $F_2$ are $G$-invariant by 
replacing $F_i$ by 
\[ \tilde{F_i}=\int_G F_i d\mu,\] 
where $\mu$ is the Haar measure on $G$, so that $\int_G 1d \mu=1$. 
We set 
\[ \pb{f_1}{f_2}_N\equiv\imath^*\pb{F_1}{F_2}\] 
and show that this definition is independent of the choice of 
$F_i$. For this we observe that, since the action of $G$ is Poisson, 
the Hamiltonian vector field $X_F$ associated with a $G$-invariant 
function $F:M\to\R$ is tangent to $N$. It follows that the Poisson 
bracket $\pb{F_1}{F_2}|_N$, where $F_1$ and $F_2$ are $G$-invariant 
functions, depends only on the restrictions ${F_i}|_N$. 
 
It is obvious from its definition that $\Pb_N$ is 
bilinear, and satisfies the Jacobi and Leibniz identities, for 
these identities hold for $\Pb$. Hence, we obtain a Poisson 
bracket on $N$, and it is the only Poisson bracket satisfying 
(\ref{eq:invol:poisson}). 
\qed 
\begin{rem} 
The previous result can be seen as a particular case of Dirac 
reduction (for the general theorem on Dirac reduction, see 
Weinstein (\cite{Wein1}, Prop.~1.4) and Courant (\cite{Cour}, 
Thm.~3.2.1). The case where $G$ is a reductive algebraic group is 
discussed in \cite{Van1} using a different method. 
\end{rem} 

Since this result applies in particular when $G$ is a finite group, we have: 
 
\begin{cor} 
  Suppose that $(M,\Pb)$ is a Poisson manifold, and $G$ is a finite group 
  acting on $M$ by Poisson automorphisms. Then the fixed point set $N=M^G$ 
  carries a (unique) Poisson structure $\Pb_N$ satisfying equation 
  (\ref{eq:invol:poisson}). 
\end{cor} 
 
Let us consider the special case $G=\Z_2$. Then $G=\set{I,\phi}$, 
where $\phi:M\to M$ is a Poisson involution. We conclude that 
$N=M^G=\set{x:\phi(x)=x}$ has a unique Poisson bracket satisfying 
equation (\ref{eq:invol:poisson}).  So we see that Theorem 
\ref{thm:invol:poisson} contains as a special case the following 
result, which is known as the \emph{Poisson involution theorem} (see 
\cite{Fer3,Xu}). 
 
\begin{cor} 
\label{cor:Poisson:involution} 
  Suppose that $(M,\Pb)$ is a Poisson manifold, and $\phi:M\to M$ is a 
  Poisson involution. Then the fixed point set $N=\set{x\in M:\phi(x)=x}$ 
  carries a (unique) Poisson structure $\Pb_N$ such that 
  \[ 
    \imath^*\pb{F_1}{F_2}=\pb{\imath^*F_1}{\imath^*F_2}_N 
  \] 
  for all functions $F_1,F_2\in C^\infty(M)$ invariant under $\phi$. 
\end{cor}

One should note that, in general, the fixed point set $M^G$ is not 
a Poisson submanifold of $M$, since it is not a union of symplectic 
leaves of $M$.  In other words, underlying Theorem 
\ref{thm:invol:poisson} there is a true Dirac type reduction, 
rather than a restriction or Poisson reduction.

%
\section{The Toda lattices}            
%
\label{Toda} 
\subsection{The $A_n$-Toda system} 
 
The phase space $\T_n$ of the (non-periodic) $A_{n-1}$-Toda lattice is 
the affine variety of all Lax operators in $\sllie_n$ of the 
form

\begin{equation} 
\label{Toda:Lax:operator} 
  L=\left( 
  \begin{array}{cccccc} 
    b_1&a_1& 0 &\cdots& 0 & 0\\ 
     1 &b_2&a_2&      &   & 0\\ 
     0 & 1 &   &      &   &\vdots\\ 
    \vdots&&\ddots&\ddots& &\vdots\\ 
    0 &  & & & b_{n-1} &a_{n-1}\\ 
    0 &0&\cdots &\cdots&1&b_n 
  \end{array} 
  \right). 
\end{equation} 
The reader will notice that we adopt a set of variables (due to 
Kostant) that differ slightly from the usual Flaschka variables 
(see \cite{Aud}, Chp.~V). This Lax pair has an obvious Lie 
algebraic interpretation which leads to the generalized Toda 
Lattices (see next paragraph).

The Hamiltonians $H_k$, $(k=1,\dots,n)$, defined by 
\begin{displaymath} 
  H_k=\frac{1}{k}\tr(L^{k}). 
\end{displaymath} 
are in involution with respect to the linear Poisson bracket 
$\Pbt^1$, defined by 
\begin{align} 
\label{eq:linear:bracket} 
    \pbt{a_i}{a_j}^1&=\pbt{b_i}{b_j}^1=0\\ 
    \pbt{a_i}{b_j}^1&=a_i(\delta_{ij}-\delta_{i+1,j}).\notag 
\end{align} 
The commuting vector fields $X_k=\pbt{\cdot\,}{H_k}^1$ admit the Lax 
representation 
\begin{equation} 
\label{Toda:Lax} 
  X_k(L)=[L,(L^{k+1})_+], 
\end{equation} 
where the subscript $+$ denotes projection into the Lie subalgebra 
of $\sllie_n$ generated by the positive roots. In \cite{Mos}, it 
was proved that the non-periodic Toda lattice is a completely 
integrable system, by applying a finite dimensional analogue of the 
inverse scattering method. For the periodic case, the flows are 
linear on the generic fibers of the momentum map $K:\T_n\to\C[x]$, 
and since these fibers are affine parts of hyperelliptic Jacobians, 
the periodic Toda lattice is an a.c.i.\ system (see \cite{AvM} for 
details). 
 
The Toda lattice is also Hamiltonian with respect to a quadratic 
Poisson bracket $\Pbt^2$, defined by 
\begin{align} 
\label{eq:quadratic:bracket} 
    \pbt{a_i}{a_j}^2&=a_ia_j(\delta_{i,j+1}-\delta_{i+1,j}), 
  \qquad \pbt{a_i}{b_j}^2=a_ib_j(\delta_{i,j}-\delta_{i+1,j}),\\ 
    \pbt{b_i}{b_j}^2&=a_i(\delta_{i,j+1}-\delta_{i+1,j}).\notag 
\end{align} 

The quadratic Toda bracket appeared in a paper of Adler \cite{adler} in 1979.
The brackets $\Pbt^1$ and $\Pbt^2$ are compatible, and are in fact part of a 
full hierarchy of higher order Poisson brackets \cite{Dam1,Fer1}. Denote 
by $\pi_{1}$ and $\pi_{2}$ the Poisson tensors associated with the 
linear and quadratic brackets. Also, let $Z_{0}$ be the Euler vector field 
\begin{equation} 
\label{eq:euler} 
  Z_0=\sum_{i=1}^n a_i\frac{\partial}{\partial a_i}+ 
  \sum_{i=1}^n b_i\frac{\partial}{\partial b_i}. 
\end{equation} 
The following result is proved in \cite{Fer1}: 
\begin{prop} 
There exists a sequence of Poisson tensors $\pi_k$ and master 
symmetries $Z_k$, $k=0,1,2,\dots$ such that 
\begin{enumerate} 
\item[(i)] The $\pi_k$ are compatible Poisson tensors and the 
functions $H_k$ are in involution with respect to all of the $\pi_k$; 
\item[(ii)] The vector fields $X_k$ admit the multiple Hamiltonian 
formulation 
\[ X_{k+l-1}=\pi_k dH_l=\pi_{k-1}dH_{l+1};\] 
\item[(iii)] The Poisson tensors, the integrals and the Hamiltonian 
vector fields, satisfy the deformation relations: 
\begin{align} 
\label{eq:deformation} 
  \Lie_{Z_k}\pi_l&=(l-k-2)\pi_{k+l},& Z_k(H_l)&=(k+l)H_{k+l},\\ 
  \qquad \Lie_{Z_k}X_l&=(l-1)X_{k+l},& [Z_k,Z_l]&=(k-l)Z_{k+l}.\notag 
\end{align} 
\end{enumerate} 
\end{prop} 

These relations will be important later to deduce some properties 
of the Volterra lattices. Using  these relations  we can find explicit formulas 
for the higher order Poisson brackets. For example, we have that 
\begin{multline} 
\label{eq:master:symmetry} 
  Z_1=\sum_{i=1}^{n-1} 
  a_i\left(b_i(1-2i)+b_{i+1}(1+2i)\right)\frac{\partial}{\partial 
  a_i}\\ 
  +\sum_{i=1}^{n} 
  \left(a_{i-1}(2-2i)+a_{i}(2+2i)+b_i^2\right)\frac{\partial}{\partial 
  b_i} 
\end{multline} 
so we find that the cubic bracket $\pi_3=-\Lie_{Z_1}\pi_2$ is given 
by (see \cite{Dam3}): 
\begin{align} 
\label{eq:cubic:bracket} 
    \pbt{a_i}{a_{i+1}}^3&=2a_ia_{i+1}b_{i+1},& 
      \pbt{a_i}{b_i}^3&=-a_ib_i^2-a_i^2,\notag\\ 
      \pbt{a_i}{b_{i+1}}^3&=a_ib_{i+1}^2+a_i^2,& 
    \pbt{a_i}{b_{i+2}}^3&=a_i a_{i+1},\\ 
      \pbt{a_{i+1}}{b_i}^3&=-a_i a_{i+1},& 
      \pbt{b_i}{b_{i+1}}^3&=a_i(b_i+b_{i+1}).\notag 
\end{align} 
Similarly, one can compute any higher order Poisson bracket. 

The cubic   bracket $\pi_3$
 was found by Kupershmidt \cite{kup} via the infinite Toda lattice.

\subsection{The $B_{n}$-Toda system} 
\label{section:B:Toda} 
Generalizations of the Toda lattice for other root systems are well 
known and were first given in \cite{Bog1}. We consider here only the case 
of $B_n$. 
 
The phase space for the $B_n$-Toda lattice is the affine 
variety of all Lax operators in $\sllie_{2n+1}$ of the form: 
\begin{equation} 
\label{Toda:B:Lax:operator} 
  L=\left( 
  \begin{array}{cccccccccccc} 
    b_1&a_1& 0 &\cdots& &   & &  &\cdots& 0 & 0\\ 
     1 &b_2&a_2&      &  &  & &  &      &   & 0\\ 
    \vdots&  &\ddots&\ddots & &&  & &     &   &\vdots\\ 
     &  &  & b_{n-1} &a_{n-1}\\ 
     & &   &1&b_n & a_n \\ 
     & &   & &1 & 0 & -a_n\\ 
     & &   & &  & -1 & -b_n &-a_{n-1}\\ 
     & &   & &  &    & -1&-b_{n-1}\\ 
     \vdots& &   &   &    &   &&\ddots &\ddots&&\vdots\\ 
     0& &   && &  &       &&-1&-b_2&-a_1 \\ 
     0&0 &\cdots &  & &      &   &&0&-1&-b_1 
  \end{array} 
  \right) 
\end{equation} 
This matrix actually lies in the real simple Lie algebra 
$\mathfrak{g=so}(n,n+1)$. We would like to have a multiple 
Hamiltonian formulation for the $B_n$-Toda systems 
\begin{equation} 
\label{B:Toda:Lax} 
  \frac{dL}{dt}=[L,(L^{k+1})_+]. 
\end{equation} 
Note that, in this case, only the even powers of $L$ give 
non-trivial integrals 
\begin{displaymath} 
  H_{2k}=\frac{1}{2k}\tr(L^{2k}),\qquad (k=1,2,\dots). 
\end{displaymath} 

In order to write down a Hamiltonian formulation for these systems 
we recall that the $B_n$-Toda is a subsystem of the 
$A_{2n}$-system. This is well known and in fact, more is true: we 
can find a symmetry such that the $B_n$-Toda is obtained  through a 
symmetry reduction from the $A_{2n}$-Toda. To see this, observe 
that conjugation by a diagonal matrix takes the Lax matrix 
(\ref{Toda:B:Lax:operator}) to the Lax matrix: 
\begin{equation} 
\label{Toda:B:Lax:operator:other} 
  L=\left( 
  \begin{array}{cccccccccccc} 
    b_1&a_1& 0 &\cdots& &   & &  &\cdots& 0 & 0\\ 
     1 &b_2&a_2&      &  &  & &  &      &   & 0\\ 
    \vdots&  &\ddots&\ddots & &&  & &     &   &\vdots\\ 
     &  &  & b_{n-1} &a_{n-1}\\ 
     & &   &1&b_n & a_n \\ 
     & &   & &1 & 0 & a_n\\ 
     & &   & &  & 1 & -b_n &a_{n-1}\\ 
     & &   & &  &    & 1&-b_{n-1}\\ 
     \vdots& &   &   &    &   &&\ddots &\ddots&&\vdots\\ 
     0& &   && &  &       &&1&-b_2&a_1 \\ 
     0&0 &\cdots &  & &      &   &&0&1&-b_1 
  \end{array} 
  \right) 
\end{equation} 
and this is just a matrix in $\T_{2n+1}$ satisfying 
$a_{2n+1-i}=a_i$ and $b_{2n+2-i}=-b_i$. Therefore, if we consider 
the involution $\phi:\T_{2n+1}\to\T_{2n+1}$ defined by 
\[ \phi(a_i,b_i)=(a_{2n+1-i},-b_{2n+2-i}),\] 
then the phase space of the $B_n$-Toda can be identified with the 
fixed point set of this involution. On the other hand the 
$\Z_2$-action generated by this involution is a Poisson action for 
the odd Poisson brackets. More precisely, we have: 
\begin{prop} 
\label{prop:B:Toda} 
The map $\phi:\T_{2n+1}\to\T_{2n+1}$ satisfies 
\[\phi_*\pi_k=(-1)^{k+1}\pi_k.\] 
\end{prop} 
\begin{proof} 
For lower order Poisson brackets one can check this relation by direct 
computation. For higher Poisson brackets, we note that expression 
(\ref{eq:master:symmetry}) for the master symmetry $Z_1$, shows after 
some tedious computation that 
\[ \phi_* Z_1=-Z_1+X,\] 
where $X$ is a multiple of the first Hamiltonian vector field 
$X_1$. Now relation (\ref{eq:deformation}) gives 
\[ \Lie_{Z_1}\pi_k=(k-3)\pi_{k+1},\] 
so it follows by induction that 
\begin{align*} 
  \phi_*\pi_{k+1}&=\frac{1}{k-3}\phi_* \Lie_{Z_1}\pi_k\\ 
                 &=\frac{1}{k-3}\Lie_{\phi_*Z_1}\phi_*\pi_k\\ 
                 &=\frac{1}{k-3}\Lie_{-Z_1} 
                 (-1)^{k+1}\pi_k=(-1)^{k+2}\pi_{k+1}. 
\end{align*} 
\end{proof} 

Using Corollary \ref{cor:Poisson:involution}, we conclude that the 
odd Poisson brackets for the $A_{2n}$-Toda induce Poisson brackets 
for the $B_n$-Toda lattices, which therefore possesses a multiple 
Hamiltonian formulation. In this way we have deduced the following 
result of \cite{Dam3}: 
\begin{cor} 
There exists a sequence of Poisson tensors 
$\pi_1,\pi_3,\pi_5,\dots$ such that the $B_n$-Toda lattices 
(\ref{B:Toda:Lax}) possess a multiple Hamiltonian formulation: 
\[ \pi_{k+2} dH_{2l}=\pi_k dH_{2l+2}.\] 
\end{cor} 

It should be noted that Theorem \ref{thm:invol:poisson} gives an 
effective procedure to compute the Poisson brackets, as is illustrated by 
the following example. 
 
\begin{ex} 
\label{ex:cubic:B:Toda} 
In this example we compute the cubic $B_n$-bracket.  We denote by 
$(a_i,b_j)$, where $i,j=1,\dots,n$, the coordinates on the phase space 
of the $B_n$-Toda lattice, and by $(\tilde{a}_i,\tilde{b}_{j})$, where 
$i=1,\dots,2n$, $j=1,\dots,2n+1$, the coordinates on the phase space 
of the $A_{2n}$-Toda lattice. 
 
Suppose we would like to compute the cubic $B_n$-bracket 
$\pbt{a_i}{b_i}^3$. First we extend the function $a_i$ and $b_i$ to 
invariant functions $\widehat{a}_i$ and $\widehat{b}_i$ on the phase space of 
the $A_{2n}$-Toda lattice. For example, we can take the functions 
\[ \widehat{a}_i=\frac{\tilde{a}_i+\tilde{a}_{2n+1-i}}{2},\qquad 
\widehat{b}_i=\frac{\tilde{b}_i-\tilde{b}_{2n+2-i}}{2}.\] 
Then we compute the $A_{2n}$-cubic bracket of these functions: we find 
using (\ref{eq:cubic:bracket}) that for $i<n$, we have 
\[ 
\pbt{\widehat{a}_i}{\widehat{b}_i}^3=-\frac{1}{4} 
\left(\tilde{a}_i\tilde{b}_i^2+\tilde{a}_i^2+ 
\tilde{a}_{2n+1-i}\tilde{b}_{2n+2-i}^2+\tilde{a}_{2n+1-i}^2\right), 
\] 
while for $i=n$ we obtain 
\[ 
\pbt{\widehat{a}_n}{\widehat{b}_n}^3=-\frac{1}{4} 
\left(\tilde{a}_n\tilde{b}_n^2+\tilde{a}_n^2+ 
\tilde{a}_{n+1}\tilde{b}_{n+2}^2+\tilde{a}_{n+1}^2 
+2\tilde{a}_n\tilde{a}_{n+1} 
\right). 
\] 
Finally, we restrict the brackets to the $B_n$-phase space, to obtain 
the cubic $B_n$-bracket 
\begin{align} 
\label{eq:B:cubic:bracket:1} 
\pbt{a_i}{b_i}^3&=-\frac{1}{2}\left(a_i b_i^2+a_i^2\right),\qquad (i<n)\\ 
\pbt{a_n}{b_n}^3&=-\frac{1}{2}\left(a_n b_n^2+2a_n^2\right). 
\end{align} 
The other brackets are computed in a similar fashion and they give 
the following expressions 
\begin{align} 
\label{eq:B:cubic:bracket:2} 
    \pbt{a_i}{a_{i+1}}^3&=a_ia_{i+1}b_{i+1},& 
      \pbt{b_i}{b_{i+1}}^3&=a_i(b_i+b_{i+1}),\notag\\ 
      \pbt{a_i}{b_{i+1}}^3&=\frac{1}{2}\left(a_ib_{i+1}^2+a_i^2\right),& 
    \pbt{a_i}{b_{i+2}}^3&=\frac{1}{2}a_i a_{i+1},\\ 
      \pbt{a_{i+1}}{b_i}^3&=-\frac{1}{2}a_i a_{i+1}.\notag 
\end{align} 
\end{ex} 
 
\begin{rem} 
It is not hard to check that the (non-periodic) $C_n$-Toda system 
is a subsystem of the $A_{2n-1}$-Toda system. The reader can check 
that the multiple Hamiltonian structure for the $C_n$-Toda lattices 
found in \cite{Dam4}, \cite{DN} can be obtained from the multiple 
Hamiltonian structure for the $A_{2n-1}$-Toda systems, as in the 
$B_n$-case. In fact, the involution $\phi:\T_{2n}\to\T_{2n}$ 
defined by 
\[ \phi(a_i,b_i)=(a_{2n+1-i},-b_{2n+2-i}).\] 
satisfies $\phi_*\pi_k=(-1)^k\pi_k$, so the same method applies. 
\end{rem} 
%
\section{The Volterra Lattices}            
%
\label{Volterra} 
\subsection{The $A_n$-Volterra system} 
 
We now turn to the (non-periodic) $A_n$-Volterra system.  Its phase 
space $\K_n$ is the subspace of $\T_n$ consisting of all Lax operators 
(\ref{Toda:Lax:operator}) with zeros on the diagonal:%
\begin{equation} 
\label{Volterra:Lax:operator} 
  L=\left( 
  \begin{array}{cccccc} 
     0&a_1& 0 &\cdots& 0 & 0\\ 
     1 &0 &a_2&      &   & 0\\ 
     0 & 1 &   &      &   &\vdots\\ 
    \vdots&&\ddots&\ddots& &\vdots\\ 
    0 &  & & & 0 &a_{n-1}\\ 
    0 &0&\cdots &\cdots&1&0 
  \end{array} 
  \right). 
\end{equation} 
 Note that $\K_n$ is not a Poisson 
subspace of $\T_n$.  However, $\K_n$ is the fixed manifold of the 
involution $\psi:\T_n\to \T_n$ defined by 
\[((a_1,a_2\dots,a_n),(b_1,b_2\dots,b_n))\mapsto 
((a_1,a_2\dots,a_n),(-b_1,-b_2\dots,-b_n)),\] and we have (see 
\cite{Fer3}): 
 
\begin{prop} 
\label{prop:A:Volterra} 
$\psi:\T_n\to \T_n$ is a Poisson automorphism of $(\T_n,\Pbt^k)$, 
if $k$ even. 
\end{prop} 
 
\begin{proof} 
Again, one checks by direct computation for $k<4$ that 
\[ \psi_*\pi_k=(-1)^k\pi_k.\] 
On the other hand, we have that $\psi_*Z_1=-Z_1$, so by induction 
we see that 
\begin{align*} 
  \psi_*\pi_{k+1}&=\frac{1}{k-3}\psi_* \Lie_{Z_1}\pi_k\\ 
                 &=\frac{1}{k-3}\Lie_{\psi_*Z_1}\psi_*\pi_k\\ 
                 &=\frac{1}{k-3}\Lie_{-Z_1} 
                 (-1)^k\pi_k=(-1)^{k+1}\pi_{k+1}. 
\end{align*} 
and the result follows. 
\end{proof} 
 
Therefore, by Corollary \ref{cor:Poisson:involution}, $\K_n$ 
inherits a family of Poisson brackets $\pi_2,\pi_4,\dots$. For example, 
the quadratic bracket can be computed from formulas 
(\ref{eq:quadratic:bracket}), in a form entirely analogous to Example 
\ref{ex:cubic:B:Toda}, and is given by 
\begin{equation} 
\label{eq:quadratic:Volterra} 
  \pbk{a_i}{a_j}^2=a_ia_j(\delta_{i,j+1}-\delta_{i+1,j}), 
\end{equation} 
while the Poisson bracket $\pi_4$ is found to be given by the 
formulas 
\begin{align} 
\label{eq:cubic:Volterra} 
  \pbk{a_i}{a_{i+1}}^4&=a_ia_{i+1}(a_i+a_{i+1}),& (i&=1,\dots,n-1)\\ 
  \pbk{a_i}{a_{i+2}}^4&=a_ia_{i+1}a_{i+2},& (i&=1,\dots,n-2).\notag 
\end{align} 
The formulas for the brackets $\pi_2$ and $\pi_4$ appeared in 
\cite{Dam2}, and the existence of this hierarchy of Poisson 
brackets is proved in \cite{Fer3}. 
The first three  Poisson structures of  this system appeared first in \cite%
{Fadeev} where the Volterra lattice is treated in detail,
however the quadratic and cubic brackets do not coincide with the brackets in \cite{Dam2}.
It follows that the restriction of the integrals $H_{2k}$ to $\K_n$ 
gives a set of commuting integrals, with respect to these Poisson 
brackets. Also, for $i$ odd, the Lax equations (\ref{Toda:Lax}) 
lead to Lax equations for the corresponding flows, merely by 
putting all $b_i$ equal to zero. Taking $i=1$, we recover the 
system 
\begin{equation}\label{KM} 
  \dot{a}_i=a_i(a_{i-1}-a_{i+1}),\qquad i=1,\dots,n, 
\end{equation} 
which we called the $A_n$-Volterra lattice in the introduction. 
More generally, taking $i$ odd we find a family of commuting 
Hamiltonian vector fields on $\K_n$ which are restrictions of the 
Toda vector fields, while for $i$ even the Toda vector fields $X_i$ 
are not tangent to $\K_n$. Hence, we obtain a family of integrable 
systems admitting a multiple Hamiltonian formulation: 
\[ X_{2(k+l)-1}=\pi_{2k+2} dH_{2l}=\pi_{2k} dH_{2l+2}, \qquad (k=1,2,\dots).\] 
This system was shown to be a completely integrable system in 
\cite{Mos}, while the periodic version of this system was proved to 
be an a.c.i.\ system in \cite{Fer2,Fer3}.

\subsection{Generalized Volterra systems} 
We now describe the construction of the generalized Volterra 
systems of Bogoyavlensky (see \cite{Bog2}). 
 
Let $\gg$ be a simple Lie algebra, with $\rank\gg=n$, and let 
$\Pi=\set{\omega_{1},\omega_{2},\ldots,\omega_{n}}$ be a 
Cartan-Weyl basis for the simple roots in $\gg$. There exist unique 
positive integers $k_{i}$ such that 
\[k_{0}\omega_{0}+k_{1}\omega_{1}+\cdots+k_{n}\omega_{n}=0,\] 
where $k_{0}=1$ and $\omega _{0}$ is the minimal negative root. We 
consider the Lax pair: 
\[\dot{L}=\left[B,L\right],\] 
where 
\begin{align*} 
L(t)=&\sum_{i=1}^{n} b_{i}(t)e_{\omega_{i}}+e_{\omega_{0}} 
+\sum_{1\leq i<j\leq n} [e_{\omega_{i}},e_{\omega_{j}}],\\ 
B(t)=&\sum_{i=1}^{n} \frac{k_{i}}{b_{i}(t)}e_{-\omega_{i}} 
+e_{-\omega_{0}}. 
\end{align*} 
 
Let $\hh\subset\gg$ be the Cartan subalgebra. For every root 
$\omega_{a}\in\hh^*$ there is a unique $H_{\omega_{a}}\in\hh$ such 
that $\omega(h)=\beta\left(H_{\omega_{a}},h\right)$, for all 
$h\in\hh$, where $\beta$ denotes the Killing form. Also, $\beta$ 
induces an inner product on $\hh^*$ by setting 
$\langle\omega_{a},\omega_{b}\rangle= 
\beta\left(H_{\omega_{a}},H_{\omega_{b}}\right)$, and we define 
\[ 
c_{ij}=\left\{ 
\begin{array}{cc} 
1 & \text{if }\left\langle \omega _{i},\omega _{j}\right\rangle \neq 0\text{ 
and }i<j \\ 
0 & \text{if }\left\langle \omega _{i},\omega _{j}\right\rangle =0\text{ or }%
i=j\text{ \ } \\ 
-1\text{ } & \text{if }\left\langle \omega _{i},\omega _{j}\right\rangle 
\neq 0\text{ and }i>j%
\end{array}%
\right. 
\] 
With these choices, the Lax pair above is equivalent to the system 
of o.d.e.'s 
\begin{equation} 
\label{a1} 
\dot{b}_{i}=-\sum_{j=1}^{n}\frac{k_{j}c_{ij}}{b_{j}}. 
\end{equation} 
 
To obtain a Lotka-Volterra type system one introduces a new set of 
variables by 
\begin{align*} 
x_{ij}&= c_{ij}b_{i}^{-1}b_{j}^{-1},\\ x_{ji}&= -x_{ij},\\ x_{jj}&= 
0. 
\end{align*} 
Note that $x_{ij}\not=0$ iff there exists an edge in the Dynkin 
diagram for the Lie algebra $\gg$ connecting the vertices 
$\omega_{i}$ and $\omega_{j}$. System (\ref{a1}), in the variables 
$x_{ij}$, takes the form 
\begin{equation} 
\label{a2} 
\dot{x}_{ij}=x_{ij}\sum_{s=1}^{n}k_{s}\left(x_{is}+x_{js}\right), 
\end{equation} 
which is a Lotka-Volterra type system. We call (\ref{a2}) the 
Bogoyavlensky-Volterra system associated with $\gg$ (or the 
$\gg$-Volterra system for short). 
 
\begin{ex} 
Let us take $\gg=A_n$. Then we have the Dynkin diagram 
\[ 
\xymatrix{\bullet \ar@{-}[r]&\bullet \ar@{-}[r]&\bullet \ar@{..}[r]& 
\bullet\ar@{-}[r]&\bullet\ar@{-}[r]&\bullet} 
\] 
with $k_i=1$, $i=0,\dots,n$. If we label the edges of this diagram 
by $a_1,\dots,a_{n}$, then system (\ref{a2}) takes precisely the 
form of the $A_n$-Volterra lattice. 
\end{ex} 
 
We now  turn our attention to the $B_n$-case. 
 
\subsection{The $B_{n}$-Volterra system} 
For $\gg=B_n$, the Dynkin diagram has the form 
\[ 
\xymatrix{\bullet \ar@{-}[r]&\bullet \ar@{-}[r]&\bullet \ar@{..}[r]& 
\bullet\ar@{-}[r]&\bullet\ar@{=}[r]&\bullet} 
\] 
and for the obvious labeling (and after a linear change of 
variables) we obtain the $B_n$-Volterra system 
\begin{equation} 
\label{a3} 
\left\{ 
\begin{array}{l} 
\dot{a}_{1}=-a_{1}a_{2} \\ 
\dot{a}_{i}=a_{i}(a_{i-1}-a_{i+1}),\qquad (i=2,\dots, n-1)\\ 
\dot{a}_{n}=a_n\left(a_{n-1}+a_n\right). 
\end{array} 
\right. 
\end{equation} 
 
The multiple Hamiltonian structure for the $B_n$-Volterra system 
can be obtained from the $A_{2n}$-Volterra system in the same 
fashion as the multiple Hamiltonian formulation for the $B_n$-Toda 
system was obtained from the $A_{2n}$-Toda system (cf.~Section 
\ref{section:B:Toda}). In fact, we notice that equations (\ref{a3}) 
for the $B_n$-Volterra system can be obtained from equations 
(\ref{KM}) for the $A_{2n}$-Volterra system by setting 
$a_{2n+i-i}=-a_i$ (this defines an invariant submanifold). Hence, 
we take as phase space of the $B_n$-Volterra system the Lax 
operators of the form (\ref{Volterra:Lax:operator}) satisfying 
these additional restrictions: 
\begin{equation} 
\label{Volterra:B:Lax:operator} 
  L=\left( 
  \begin{array}{cccccccccccc} 
     0&a_1& 0 &\cdots& &   & &  &\cdots& 0 & 0\\ 
     1 &0&a_2&      &  &  & &  &      &   & 0\\ 
    \vdots&  &\ddots&\ddots & &&  & &     &   &\vdots\\ 
     &  &  & 0&a_{n-1}\\ 
     & &   &1&0 & a_n \\ 
     & &   & &1 & 0 & -a_n\\ 
     & &   & &  & 1 & 0 &-a_{n-1}\\ 
     & &   & &  &    & 1&0\\ 
     \vdots& &   &   &    &   &&\ddots &\ddots&&\vdots\\ 
     0& &   && &  &       &&1&0&-a_1 \\ 
     0&0 &\cdots &  & &      &   &&0&1&0 
  \end{array} 
  \right), 
\end{equation} 
Notice that this subspace appears as the fixed point set of the 
involution $\varphi:\K_{2n}\to\K_{2n}$ defined by 
\[ \varphi(a_i)=-a_{2n+1-i}.\] 
We will see below that this involution can be used to obtain the 
multiple Hamiltonian structure of the $B_n$-Volterra system. 
 
\begin{rem} 
There is a diagonal matrix $D$ which conjugates the Lax matrix 
given above for the $B_n$ system, to the following Lax matrix 
\begin{equation} 
\label{Volterra:B:Lax:operator:another} 
  L=\left( 
  \begin{array}{cccccccccccc} 
     0&a_1& 0 &\cdots& &   & &  &\cdots& 0 & 0\\ 
     1 &0&a_2&      &  &  & &  &      &   & 0\\ 
    \vdots&  &\ddots&\ddots & &&  & &     &   &\vdots\\ 
     &  &  & 0&a_{n-1}\\ 
     & &   &1&0 & a_n \\ 
     & &   & &1 & 0 & a_n\\ 
     & &   & &  & -1 & 0 &a_{n-1}\\ 
     & &   & &  &    & -1&0\\ 
     \vdots& &   &   &    &   &&\ddots &\ddots&&\vdots\\ 
     0& &   && &  &       &&-1&0&a_1 \\ 
     0&0 &\cdots &  & &      &   &&0&-1&0 
  \end{array} 
  \right). 
\end{equation} 
Notice that these Lax matrices \emph{are not} obtained as Lax 
matrices for the $B_n$-Volterra system (matrices of the form 
(\ref{Toda:B:Lax:operator})) with zeros on the diagonal. 
\end{rem} 
 
We now have the following proposition analogous to Propositions 
\ref{prop:B:Toda} and \ref{prop:A:Volterra}. The proof is similar 
and will be omitted. 
 
\begin{prop} 
Let $\pi_k$ ($k=2,4,\dots$) be the Poisson tensors of the 
$A_{2n}$-Volterra system. Then 
\[ \varphi_*\pi_k=(-1)^{k/2}\pi_k.\] 
\end{prop} 
 
In this way, by Corollary \ref{cor:Poisson:involution}, the phase 
space of the $B_n$-Volterra system carries Poisson brackets $\pi_4, 
\pi_8,\dots$. For example, from (\ref{eq:cubic:Volterra}) we can 
compute, using the same technique as in Example 
\ref{ex:cubic:B:Toda}, the following explicit formulas for the 
bracket $\pi_4$: 
\begin{align} 
\label{eq:cubic:B:Volterra} 
  \pbk{a_i}{a_{i+1}}^4&=\frac{1}{2}a_ia_{i+1}(a_i+a_{i+1}),\notag\\ 
  \pbk{a_{n-1}}{a_n}^4&=\frac{1}{2}a_{n-1}a_n(a_{n-1}+2a_n),\qquad (i=1,\dots,n-2)\\ 
  \pbk{a_i}{a_{i+2}}^4&=\frac{1}{2}a_ia_{i+1}a_{i+2}.\notag 
\end{align} 
The restriction of the invariant functions 
\[ H_{4k}=\frac{1}{4k}\tr(L^{4k})\] 
to this space, defines a hierarchy of systems, possessing a 
multiple Hamiltonian formulation: 
\[ X_{4(k+l)-1}=\pi_{4k+4} dH_{4l}=\pi_{4k} dH_{4l+4}, \qquad (k=1,2,\dots).\] 
The function $I_4=\frac{1}{4}\sum_{i=1}^{n-1} (2a_i^2+a_i a_{i+1})$ 
gives the $B_n$-Volterra system (\ref{a3}). The brackets above, as 
well as the master symmetries for these systems (including the 
$C_n$ and $D_n$ cases) appear in a recent preprint of Kouzaris 
(\cite{Kou}). However, the Lax matrices above lead to Lax pairs 
which are \emph{different} from the Lax pairs given by Kouzaris. 
 
Finally we remark that the $C_n$-Volterra system can be identified 
with the $B_n$-Volterra systems (see \cite{Kou}), and hence admits 
the same description. 
 
\section{Epilogue: From Toda to Volterra and back}           %
%
\label{Toda:Volterra} 

In this concluding section we would like to explain what we know so 
far about the connection between the Toda and Volterra systems, 
including the results obtained above. The relation between the two 
systems relies on special symmetries of the phase spaces. 
 
We have shown above that the phase space of the $A_n$-Volterra 
system appears as the fixed point set of a Poisson involution of 
the $A_n$-Toda phase space. On the other hand, we have also shown 
that the multiple Hamiltonian structure of the $B_n$-Toda system 
can be obtained from the $A_{2n}$-Toda system. Hence, to get to the 
$B_n$-Volterra system, there are \emph{a priori} two distinct ways 
to proceed, as explained by the following diagram: 
\smallskip 
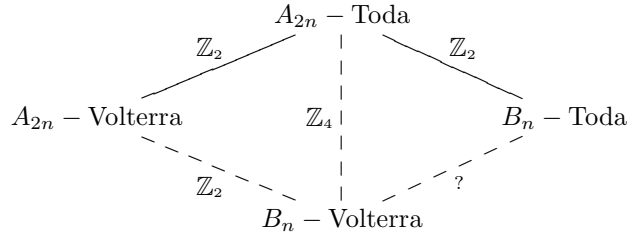
\[ 
\xymatrix{ 
&A_{2n}-\text{Toda}\ar@{-}[dl]_{\Z_2}\ar@{-}[dr]^{\Z_2}\ar@{--}[dd]_{\Z_4}\\ 
A_{2n}-\text{Volterra}\ar@{--}[dr]_{\Z_2}& & 
B_{n}-\text{Toda}\ar@{--}[dl]^{?}\\ & B_{n}-\text{Volterra}} 
\] 
We have seen that the correct way to proceed is to choose the 
\emph{left} side of this diagram. In fact, we saw above that we can get 
from $A_{2n}$-Toda to $A_{2n}$-Volterra by a Poisson involution 
$\psi:\T_{2n}\to\T_{2n}$ (i.e., a $\Z_2$-symmetry). Also, we can 
get from $A_{2n}$-Volterra to $B_n$-Volterra using another Poisson 
involution $\varphi:\K_{2n}\to\K_{2n}$ (and, hence, again a 
$\Z_2$-symmetry). Note also that we can go straight from 
$A_{2n}$-Toda to $B_n$-Volterra using a $\Z_4$-symmetry: if one 
defines the map $\varphi:\T_{2n}\to\T_{2n}$ by 
\[ \widetilde{\varphi}(a_i,b_i)=(-a_{2n+1-i},\sqrt{-1}\ b_{2n+2-i}),\] 
then one checks that the group 
\[ 
G=\set{I,\widetilde{\varphi},\widetilde{\varphi}^2,\widetilde{\varphi}^3} 
\] 
acts by Poisson automorphisms on $(\T_{2n},\pi_{4k})$, 
$k=1,2,\dots$. The fixed point set (i.e., the phase space for 
$B_n$-Volterra) inherits Poisson structures $\pi_{4k}$ and the even 
flows reduce to this space. Notice also that 
$\widetilde{\varphi}|_{\K_{2n}}=\varphi$ and 
$\widetilde{\varphi}^2=\psi$, so this reduction in one stage (the 
middle line in the diagram) coincides with the reduction in two 
stages (the left part of the diagram). 
 
On the other hand, there seems to be no such symmetry reduction 
from the $B_n$-Toda lattice to the $B_n$-Volterra lattice: the 
various Poisson structures and Hamiltonian functions we have for 
the $B_n$-Toda systems \emph{do not} restrict to the phase space of 
the $B_n$-Volterra systems. We could however still find a 
connection between these two systems, which goes in the opposite 
direction. 
 
To explain this connection, we perform the change of variable, 
$a_i=-2 x_i^2$, and we consider the following equivalent (i.e., 
conjugate) form of the Lax matrix (\ref{Volterra:B:Lax:operator}) 
(or, equivalently, (\ref{Volterra:B:Lax:operator:another})): 
\[ 
  L=\left( 
  \begin{array}{ccccccccccc} 
    0&x_1& 0 &\cdots& &      &\cdots& 0 & 0\\ 
     x1 & 0&x_2&      &  &     &      &   & 0\\ 
    \vdots&  &\ddots&\ddots & &   &     &   &\vdots\\ 
      & &     &0 & x_n \\ 
     & &    &x_n & 0 & \sqrt{-1}x_n\\ 
     & &   &   & \sqrt{-1}x_n & 0 &\\ 
     \vdots&    &       &   &&\ddots &\ddots&&\vdots\\ 
     0& &   &&   &      &&0& \sqrt{-1}x_1 \\ 
     0&0 &\cdots    &      &   &&0&\sqrt{-1}x_1&0 
  \end{array} 
  \right). 
\] 
In these new variables the equations for the $B_n$-Volterra lattice 
(\ref{a3}) become: 
\[ 
\left\{ 
\begin{array}{l} 
\dot{x_1}=x_1 x_2^2,\\ 
\dot{x_i}= x_i (x_{i-1}^2 -x_{i+1}^2), \\ 
\dot{x_n}=-x_n (x_n^2+x_{n-1}^2 ), 
\end{array}\right. 
\qquad (i=2,\dots,n-1). 
\] 
The relation between the $B_n$-Volterra system and the 
corresponding Toda system of types $B_n$ and $C_n$ is similar to 
the relation observed by Moser (see \cite{Mos}) between the 
$A_n$-Volterra and the $A_n$-Toda lattices. One starts by taking 
the square of the Lax matrix above and notices that $L^2$ leaves 
certain subspaces invariant. Moreover, $L^2$ reduces on each of 
these invariant spaces to a symmetric Jacobi matrix. More 
precisely, assume that $L^2$ is a $N 
\times N$ matrix. Then there are two distinct cases: 
\begin{itemize} 
\item $N=4n+1:$ Removing all odd columns and all odd 
rows we end--up with an $2n \times 2n$ matrix and a Toda system of 
type $C_n$. On the other hand, removing all even columns and all 
even rows we end--up with an $2n+1 \times 2n+1$ matrix and a Toda 
system of type $B_n$. 
 
\item $N=4n+3:$ Removing all odd columns and all odd rows 
we end--up with an $2n+1 \times 2n+1$ matrix and a Toda system of 
type $B_n$. On the other hand, removing all even columns and all 
even rows we end--up with an $2 (n+1) \times 2(n+1)$ matrix and a 
Toda system of type $C_{n+1}$. 
\end{itemize} 
In this way, we have a procedure which takes us from a Volterra 
systems of type $B_n$ (or $C_n$) to either a Toda system of type 
$B_n$ or a Toda system of type $C_n$. 
 
\begin{ex} Take $N=9$ and $n=2$. Omitting all even rows and all even 
columns of $L^2$ we obtain the $5 \times 5$ matrix 
\[ 
\left( 
\begin{array}{ccccc} 
x_1^2 & x_1 x_2 & 0 & 0 & 0 \\ 
x_1 x_2 & x_2^2 +x_3^2 & x_3 x_4 & 0 & 0 \\ 
0  & x_3 x_4 & 0 & -x_3 x_4 & 0 \\ 
0 & 0 & -x_3 x_4 & -x_2^2-x_3^2 & -x_1 x_2 \\ 
0 & 0 & 0 & -x_1 x_2 & -x_1^2%
\end{array}%
\right).
\] 
We identify this  matrix with a symmetric Jacobi matrix of type 
$B_2$, i.e. we let 
\[B_1=x_1^2, B_2=x_2^2+x_3^2, A_1=x_1 x_2, A_2=x_3 x_4,\] 
and the equations satisfied by $A_1$, $A_2$, $B_1$, and $B_2$ are 
\[ 
\left\{ 
\begin{array}{l} 
\dot{A}_{1}=A_{1}\left( B_{2}-B_{1}\right), \\ 
\dot{A}_{2}= -A_2 B_2, \\ 
\dot{B}_{1}=2 A_1^2, \\ 
\dot{B}_{2}=2 A_2^2-2 A_1^2. 
\end{array} 
\right. 
\] 
These are precisely the Toda equations of type $B_2$. 
\end{ex} 
 
\bibliographystyle{amsplain}

\end{document}